%% file: template.tex
\documentclass{vgtc}                          

\graphicspath{{figures/}{pictures/}{images/}{./}} 

\usepackage{times}                     

\usepackage{tabu}                      
\usepackage{booktabs}                  
\usepackage{lipsum}                    
\usepackage{mwe}                       

\usepackage{booktabs}
\newcommand{\ra}[1]{\renewcommand{\arraystretch}{#1}}

\usepackage{soul}
\usepackage{tabularx}
\usepackage{enumitem}
\usepackage{tcolorbox}

\tcbuselibrary{breakable}

\input{commands}

\usepackage{mathptmx}                  

\onlineid{0}

\vgtccategory{Research}

\vgtcinsertpkg

\title{Context Matters! Social, Spatial, and Semantic Coordination with an Embodied Agent in Immersive Analytics}

\author{
Andrew Cunningham\thanks{e-mail: andrew.cunningham@unisa.edu.au}\\ %
        \scriptsize University of South Australia %
\and Nicolas Barbotin\thanks{e-mail:nic.barbotin@unisa.edu.au}\\ %
     \scriptsize University of South Australia %
     \and Adam Drogemuller\thanks{e-mail:adam.drogemuller@unisa.edu.au}\\ %
     \scriptsize University of South Australia %
\and Jack Fraser\thanks{e-mail:jack.fraser@unisa.edu.au}\\ %
     \scriptsize  University of South Australia
}

\teaser{
  \centering
  \includegraphics[width=\linewidth]{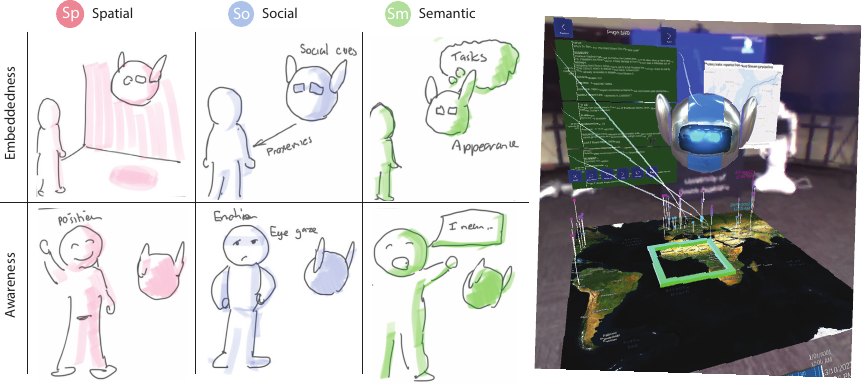}
  \caption{\textbf{(left)} Our \textit{conceptual framework}, split into dimensions of spatial, social, and semantic context. Factors within these contexts can be further categorised into embeddedness and awareness. \textbf{(right)} \textit{Cobot}, an example of an embodied AI agent in an immersive analytics scenario, demonstrating aspects of spatial, social, and semantic context in the collaboration. }
  \label{fig:teaser}
}

\abstract{
    \input{sections/abstract}
} 


\begin{document}


\firstsection{Introduction}

\maketitle

Immersive technologies such as Augmented Reality (AR) allow data visualisations to be spatially embedded in the real world, enabling users to leverage the world around them to spatially embed, visualise, and explore complex datasets.
For example, an analyst wearing an AR headset could embed a map showing world events on their office table, plot world oil prices on their wall, and create a filtered view of current events. 
This area of research, known as Immersive Analytics \cite{marriott2018immersive}, supports users in externalising their analyses and thought processes across physical space, leveraging the \textit{visual} and \textit{spatial} affordances of immersive environments. Meanwhile, advances in AI—particularly large language models (LLMs) —have enabled more natural interactions with complex systems through conversational interfaces. While most AI interactions occur through text or voice, immersive environments introduce the potential for embodied AI agents---spatially embedded avatars that represent AI systems and engage with users in the real world.

Embodied AI agents offer the potential for more ``natural'' collaboration with underlying AI systems. However, what constitutes natural interaction in the context of embodied agents within immersive environments remains poorly understood. 
Beyond the growing capabilities of LLMs, embodied agents must also meet the social and interaction expectations shaped by human-to-human collaboration. This need raises important challenges that extend beyond technical performance and require considered design work. 

In this paper, we focus on \textit{context} \cite{schilit1994context} as an integral factor in enabling natural interaction with embodied AI agents. For such agents to support seamless collaboration, they must be context-aware and able to understand and respond appropriately to the situation in which the interaction occurs. Disregarding context can disrupt the effectiveness of the user experience.

We explore the conceptual space of context-aware behaviour for embodied AI agents within immersive analytics systems, emphasising three key dimensions: \textit{spatial context} (where the collaboration takes place), \textit{social context} (the norms and dynamics of the interaction), and \textit{semantic context} (the meaning and relevance of the data and tasks at hand). We propose a preliminary framework outlining these dimensions and identify key factors within each. To ground this exploration, we present Cobot---an embodied AI agent designed for immersive analytics---as a design probe to examine how context influences interaction and to inform future design directions for embodied AI agents.

The contributions of this work are: 1) A preliminary conceptual framework of context for embodied AI agents in immersive analytics; 2) a reflective design case study of Cobot using this framework; 3) observations on design factors in context-aware embodied AI agents.

\input{sections/background}

\section{Context in Embodied Agent Interaction}

\input{contexts-table}

How we design embodied AI agents in immersive analytics systems remains an open question. The behaviours these agents should exhibit, how they look, and how they respond to the user and the environment must all be considered. 
To support these considerations, we propose that \textit{context} can be presented as a fundamental framework to help support designers when creating these agents.
We classify context for embodied AI agents in immersive analytics into three dimensions: \circlesp~\textbf{spatial context} describes how agents inhabit, move within, and respect physical space; \circleso~\textbf{social context} describes the agent adhering to social norms in interaction; and \circlesm~\textbf{semantic context} describes how agents respond meaningfully within the flow of tasks and conversation.
    
These dimensions can be further classified into \textit{embeddedness} and \textit{awareness}. Embeddedness describes the factors that situate the embodied agent in a particular dimension (e.g. rendering a shadow below the agent helps situate the agent in spatial context). Awareness refers to the factors in the dimension that the embodied agent should sense and actively reflect back to the user (e.g. the agent should be aware of the physical space and move in it appropriately to enforce spatial context).

\autoref{tab:contextual_information} presents a preliminary conceptual framework outlining key contextual dimensions and the specific factors that may shape natural interaction with an embodied agent.
In our conceptual framework, it is important that the embodied agents are designed to account for factors within these contexts; an agent should feel situated and aware of its spatial, social, and semantic context. Agents that do not consider these dimensions and their factors will create friction with the user and potentially leave them confused and frustrated (effects that we hypothesise further in \autoref{sec:usecase}).

\circlesp\ \textbf{Spatial} \textbf{embeddedness} factors emphasise illusion plausibility~\cite{slater2022}, situating the agent within the physical or virtual environment. Spatial factors such as correct scaling, occlusion, shadows, audio positioning all contribute to the illusion of plausibility. Of note, the factor of \textbf{physical coherence} \markemsp{3} suggests that the agent should act like a physical entity in the real world, having form and mass.
\circlesp\ \textbf{Spatial} \textbf{awareness} factors ensure that the agent operates and interacts with respect to the physical constraints of the space (e.g., not passing through walls or users).

\circleso\ \textbf{Social embeddedness} factors focus on social norms in interaction. These factors drawn from human interaction, such as expression and attention, reduce interaction friction and support cooperative engagement between the user and agent. 
Of note, the factor of \textbf{proxemics} \markemso{6}\ encourages designers to consider how humans use intimate, personal, and public space in social collaborations and apply those considerations to agent design.
\circleso\ \textbf{Social awareness} are the factors primarily focus on the agent interpreting and responding to the user's emotional and affective state to improve collaborations.

\circlesm\ \textbf{Semantic embeddedness} factors primarily relate to the agent \textit{appearing} appropriately in the context of the task. Of note, \textbf{Agent role} \markemsm{1} suggests that the agent should appear semantically relevant to the task. For example, a cartoon paperclip would be semantically  inappropriate to the tasks of an emergency response analytics system. 
\circlesm~\textbf{Semantic awareness} factors relate to the \textit{responding} appropriately to the task. These factors primarily relate to the information output of the system, often driven by results from the AI (e.g. large language models). 

\section{Cobot as a Design Probe}


We built Cobot, an embodied AI agent for immersive analytics, to explore several of the contextual factors identified in our preliminary conceptual framework. In this paper, we present Cobot as a design probe---being an ``object designed to ask questions and present challenges in an open-ended, often provocative manner''~\cite{Wallace2013}---into our framework. Our aim is to use Cobot to examine this conceptual space and stimulate critical reflection and discussion.

Cobot is a virtual assistant that works alongside analysts in AR to support immersive data exploration. Designed to feel like a helpful collaborator, Cobot responds to spoken commands, creates visualisations like maps or graphs in the user’s physical space, and adapts its behaviour based on the analyst's actions and surroundings. 
Rather than acting as a tool to be operated, we wanted Cobot to appear to the user as a partner, helping users make sense of complex data by maintaining spatial awareness, providing relevant visual feedback, and supporting more natural, conversational interaction.

Cobot was developed in Unity 2019.4.28f1 and ran on the Microsoft Hololens~2\footnote{https://learn.microsoft.com/en-us/hololens/}. 
Cobot was built on top of CADET~\cite{mcdade2022}, an augmented-reality based immersive analytics tool for visualising spatial and event-based data. In CADET, users can create maps, plot entities and events on those maps, read event data, create 3D graphs of entity interactions, and view timelines.

\subsection{Appearance}

Cobot was visually designed as a cartoon-like floating robot inspired by characters such as EVE from Pixar's Wall-E\footnote{https://www.pixar.com/wall-e} and Wheatley from the Valve game Portal\footnote{https://www.thinkwithportals.com/}, with eyes and ears that can emote (see \autoref{fig:teaser}, right). This aesthetic design was chosen to avoid the uncanny valley~\cite{mori2012uncanny} issue with human-like avatars. 
Cobot is visually rendered using an environment map captured from the user's environment (supporting \textbf{environment lighting}~\markemsp{1}), reflecting lights and other elements in the environment on its surface. This helps embed Cobot into the physical space, increasing the user's impression of plausibility.

\input{sections/movement}

\input{sections/usecase}

\section{Implications and Future Work}

Many of the identified features of the Cobot design probe cross-cut through contexts, implying strong relationships between certain factors within the contexts. To illustrate, consider Schmandt and Bolt's ``put-that-there'' paradigm~\cite{bolt1980put}, where a user may ask an agent to move or place something in their spaces. In this case, the embodied agent needs to  coordinate across \textbf{user gestures} \markawso{9}, \textbf{physical environment} \markawsp{4}, and \textbf{task intent} \markawsm{4}. There are likely consistent design patterns that exist across these dimensions that can assist designers in their process.

We observe, at least in our design probe, that the semantic dimension is predominantly handled by the LLM. \textbf{Agent role} \markemsm{1}, task intent \markawsm{5}, and \textbf{conversation history} \markawsm{5}, are integrated into the prompts. \textbf{Proximity to relevant visualisation}~\markemsm{2} requires some awareness of the \textbf{physical environment}~\markawsp{4}, but again, the LLM is responsible for determining what aspects of the environment and visualisation are relevant.

Informal testing suggests that Cobot’s navigation features, while simplified approximations of social \textbf{proxemics} \markemso{6}, support reasonably natural collaboration. For some factors, ``good enough'' solutions or best practices may suffice. Future work should aim to identify and evaluate these, as well as explore which contextual factors are most critical to effective interaction.

The conceptual framework is preliminary. We believe there may be more factors to identify. Our exploration has been predominantly single-user,
future iterations of the framework should examine multi-user collaboration. 


\section{Conclusion}

In this work, we identified and explored a context framework for embodied agents, encompassing spatial, social, and semantic dimensions. Coordinating these dimensions is a central concern in the design of effective embodied systems. 
Through Cobot, we demonstrate how framing design around context can be realised and address potential pain-points that effect user experience. The framework we present is preliminary; we encourage the research community to further refine and develop this theoretical framework.
Future work should investigate multi-user collaboration, examine the effects of the individual factors, and establish appropriate design patterns to help realise effective embodied AI for immersive analytics.

\acknowledgments{
The authors wish to thank Jeremy McDade for the original development work on the CADET~\cite{mcdade2022} immersive visualisation platform on which Cobot was developed.}

\bibliographystyle{abbrv-doi}

\bibliography{template}
\end{document}

%% file: commands.tex
\usepackage{xargs} 

\usepackage{xspace}
\usepackage[table]{xcolor}
\usepackage{float}
\usepackage{graphicx}
\usepackage{tikz}

\definecolor{myorange}{RGB}{249, 203, 156}
\definecolor{myblue}{RGB}{164, 194, 244}
\definecolor{mygreen}{RGB}{182, 215, 168}
\definecolor{mypurple}{RGB}{180, 167, 214}
\definecolor{myred}{RGB}{234, 153, 153}

\newcommand\circlesp{\tikz[baseline=(char.base)]{\node[circle, fill=myred, text=white, inner sep=0.4pt] (char) {\scriptsize\sffamily{Sp}};}}

\newcommand\circleso{\tikz[baseline=(char.base)]{\node[circle, fill=myblue, text=white, inner sep=0.4pt] (char) {\scriptsize\sffamily{So}};}}

\newcommand\circlesm{\tikz[baseline=(char.base)]{\node[circle, fill=mygreen, text=white, inner sep=0.4pt] (char) {\scriptsize\sffamily{Sm}};}}

\newcommand\markemsp[1]{\tikz[baseline=(char.base)]{\node[shape=rectangle, minimum size=1.2em, fill=myred, text=white, inner sep=0.4pt] (char) {\tiny\sffamily{Sp#1}};}}

\newcommand\markemso[1]{\tikz[baseline=(char.base)]{\node[shape=rectangle, minimum size=1.2em, fill=myblue, text=white, inner sep=0.4pt] (char) {\tiny\sffamily{So#1}};}}

\newcommand\markemsm[1]{\tikz[baseline=(char.base)]{\node[shape=rectangle, minimum size=1.2em, fill=mygreen, text=white, inner sep=0.4pt] (char) {\tiny\sffamily{Sm#1}};}}

\newcommand\markawsp[1]{\tikz[baseline=(char.base)]{\node[shape=circle, minimum size=1.2em, fill=myred, text=white, inner sep=0.4pt] (char) {\tiny\sffamily{Sp#1}};}}

\newcommand\markawso[1]{\tikz[baseline=(char.base)]{\node[shape=circle, minimum size=1.2em, fill=myblue, text=white, inner sep=0.4pt] (char) {\tiny\sffamily{So#1}};}}

\newcommand\markawsm[1]{\tikz[baseline=(char.base)]{\node[shape=circle, minimum size=1.2em, fill=mygreen, text=white, inner sep=0.4pt] (char) {\tiny\sffamily{Sm#1}};}}

%% file: sections/abstract.tex
Embodied AI agents in Augmented Reality (AR) could enable more natural collaboration in immersive analytics by engaging users through spatial, conversational, and visual interaction. However, designing such agents requires careful attention to context. We propose a preliminary conceptual framework for context-aware design of embodied agents structured around three dimensions of spatial, social, and semantic. This framework is further categorised into embeddedness (how agents are situated in the contexts) and awareness (how agents perceive and respond to the contexts). To explore this, we present Cobot, an embodied AI agent designed as a design probe. Cobot supports context-sensitive interaction through features such as gaze cues, spatial sound, and task-aware behaviour. We reflect on its design and identify implications for future context-aware embodied agents.

%% file: sections/background.tex
\section{Background}

In Human-Computer Interaction (HCI), context is broadly understood as the set of circumstances that shape and give meaning to interaction. Context can be broadly categorised into \textit{task}, \textit{user}, and \textit{environmental} characteristics that a system or tool will be used~\cite{maguire2001context}. Embodied interaction~\cite{dourish2001action}, which emphasises physical movement and senses in interaction as part of the cognitive process, emphasises that the context in which action occurs is critical to interpreting the meaning of that action and outcome.

\subsection{Immersive Analytics}
Immersive Analytics \cite{bach2016immersive,marriott2018immersive} can be defined as ``the use of engaging, embodied analysis tools to support data understanding and decision making''. Immersive analytics has been broadly investigated across a wide range of domains, from molecular simulation visualisation \cite{van2002virtual} to using embodied axes to construct 3D scatterplots for data analysis \cite{cordeil2017imaxes}.
Despite interest in the intersection of \textit{immersive analytics} and \textit{artificial intelligence}, the current literature exploring their intersection is sparse~\cite{wang2025immersive}. Some examples include ImmersiveIML, developed by 
Cordeil et al.~\cite{cordeil2023immersiveiml}, which enabled AI-supported user annotations of tree point cloud data in virtual reality. ImmersiveIML aimed to enable users to annotate point clouds more accurately by augmenting human capabilities with AI. Qu et al.~\cite{qu2020intelligent} investigated using 3D visualisation to visualise outputs from a machine learning model to predict whether patients had ''Rhabdomyosarcoma'', a rare soft tissue cancer. Outside of sparse examples, researchers such as Wang et al.~\cite{wang2025immersive} argue that there is a general lack of ''emerging AI techniques'' and "limited integration of AI" explored in the context of \textit{Immersive Analytics}. 

\subsection{Human-AI Collaboration}
Human-AI Collaboration~\cite{wang2020human} can be described as ``artificially intelligent agents and systems'' that aim to collaborate with humans. The \textit{collaborative} factor opposes the traditional view that artificially intelligent agents and systems are a ``black box'' with the intent of outperforming humans in benchmark tasks \cite{wang2020human}. Early demonstrations of Human-AI collaboration include Winograd's SHRDLU system~\cite{winograd1980does}. SHRDLU showcased a human conversing with a computer using text input to perform operations such as moving virtual objects and querying domains through a graphics terminal. Winograd aimed to create a system that was \textit{flexible} and \textit{adaptable} to different types of users, as opposed to adhering to a rigid command-based syntax \cite{winograd1980does}. Since the advancement and mass adoption of Large Language Models (LLMs) \cite{schillaci2024llm}, research has been increasingly investigating how AI could anthropomorphise itself as a ``collaborator'' and assistant. For example, GitHub Copilot\footnote{https://github.com/features/copilot} is a conversational agent that allows developers to ``auto-complete'' code while programming within an IDE and has seen enabled productivity gains from developers utilising the tool~\cite{ziegler2024measuring}. 

\subsection{Anthropomorphism and Agent Embodiment}
Anthropomorphism refers to ``non-human agents with human-like \textit{characteristics}, \textit{motivations}, \textit{intentions} or \textit{emotions}''~\cite{epley2007seeing}. When designing non-human agents, it is integral to consider anthropomorphic factors to make users feel more co-present with their agents \cite{nowak2003effect} and to reduce discomfort \cite{garcia2024speaking}. For example, Garcia et al. \cite{garcia2024speaking} found that in a study comparing a humanoid to a non-humanoid avatar, museum-goers preferred the non-humanoid avatar for conveying information about exhibited objects. Seminal work by Nowak and Biocca \cite{nowak2003effect} found that less human-like images elicited higher social and co-presence than human-like images. Schwind et al. \cite{schwind2018avoiding} argue for using "childish" features in avatar design, namely \textit{stylisation} (snub noses and round head shapes), to avoid uncanny valley effects. Strong atypicalities in avatar design (particularly in humanoid avatars) can elicit eeriness and promote factors such as threat avoidance \cite{chattopadhyay2016familiar}, which can create barriers to human-agent interaction.

\subsection{Context-Aware Computing}

Schilit et al.\cite{schilit1994context} describe context-aware computing as software that ``examines and reacts to an individual's changing context''. \textit{Context} relative to context-aware computing can be delineated between three aspects: 1) \textit{where you are}; 2) \textit{who you are with}; 3) \textit{what resources are nearby} \cite{schilit1994context}. Inspired by these aspects as \textit{social} and \textit{spatial} dimensions, we argue that embodied AI agents should also consider \textit{semantics} \cite{gatteschi2015semantics}. Semantics concerning AI can be described as an interpretation function between \textit{things} and \textit{symbols} \cite{zhuge2010interactive}. McCarthy \cite{mccarthy2007here} argues that to make AI \textit{human-level}, systems need to understand the ``common sense'' of an informatic situation,  reasoning and translating semantic information that could be implicit (e.g. user intention and role).   

%% file: contexts-table.tex
\begin{table*}
\caption{A conceptual framework for context in embodied AI agents}
\centering
\ra{1.3}

\small

\begin{tabularx}{\textwidth}{>{\hsize=1\hsize}p{0.25cm} *{3}{X}}
\toprule
\textbf{} & \circlesp~\textbf{Spatial Context} & \circleso~\textbf{Social Context} & \circlesm~\textbf{Semantic Context} \\
\midrule

\textit{Embeddedness}\\
& \markawsp{1}~Environment lighting and shadowing~\cite{agusanto2003photorealistic} & \markawso{1}~Expressions (e.g. thinking)~\cite{pelachaud2002subtleties,Isbister2002} & \markawsm{1}~Agent role~\cite{ISBISTER2000251} \\
& \markawsp{2}~Movement in space & \markawso{2}~Attention (through eye-gaze and body pose)~\cite{Randhavane2019} & \markawsm{2}~Proximity to relevant visualisations \\
& \markawsp{3}~Physical coherence & \markawso{3}~Vocalisation and intonation &  \\
&  & \markawso{4}~Emotion &  \\
&  & \markawso{5}~Gesturing &  \\
&  & \markawso{6}~Proxemics~\cite{mumm2011human} &  \\
\textit{Awareness}\\
& \markawsp{4}~Physical environment & \markawso{7}~User emotion & \markawsm{3}~User role \\
& \markawsp{5}~Virtual environment & \markawso{8}~User gaze & \markawsm{4}~Task intent \\
& \markawsp{6}~User position & \markawso{9}~User gestures & \markawsm{5}~Conversation history \\
&  & \markawso{10}~Turn-taking &  \\

\bottomrule
\end{tabularx}

\label{tab:contextual_information}
\end{table*}

%% file: sections/movement.tex
\subsection{Voice}

Users interact with Cobot by speaking to it. While an ideal embodied agent would continuously listen to the user and automatically identify when it should take action, we implemented a manual trigger mechanism to ease the implementation. 
Similar to existing smart assistants (e.g., Google Assistant, Siri), Cobot utilises a wake word detector \cite{albert2023conversational} to initiate listening, employing a \textit{Voice Activation Detection (VAD)} library to detect the end of a user's utterance and trigger an action.
Cobot responds with voice (\textbf{vocalisation }\markemso{3}) audio that is spatially registered. It also emits a low hum, helping users maintain awareness of its presence and position.

\subsubsection{Action inference}

To generate actions within the system from the user's voice, the system relays the user's query to OpenAI's ChatGPT (using the \texttt{gpt-3.5-turbo} model), accompanied by our custom system prompt. This system prompt describes the scene and the user's surrounding visualisations, and instructs the LLM to generate Lua code to control Cobot. This LLM model and prompting enables contextual factors of \textbf{agent role}~\markemso{1}, \textbf{task intent} ~\markemso{4}, and \textbf{conversation history} ~\markemso{5}.

\subsection{Behaviour}

Cobot's behaviour is state-based, transitioning between states based on the user's actions. Cobot's behaviour has three states: \textit{idle}, \textit{listening/thinking}, and \textit{active}.


\subsubsection{Idle}
While idle, Cobot adopts a side-by-side pose, staying on the user's right side outside their field of view (\textbf{user position} \markawsp{6}). This behaviour was adopted to feel less confrontational when not conversing with Cobot.
Cobot maintains distance from the user to avoid stepping into their personal space (tuned to 80cm) to enforce \textbf{proxemics }\markemso{6}. 
When the user turns around (rotates), Cobot moves accordingly around the user's personal space to stay on the right side of the user. This occurs after a short delay of 1 second, allowing the user to get Cobot's attention by staring at it (\textbf{user gaze }\markawso{8}) as an alternative to using the wake-word. 

\begin{figure}
    \centering
    \includegraphics[width=0.7\linewidth]{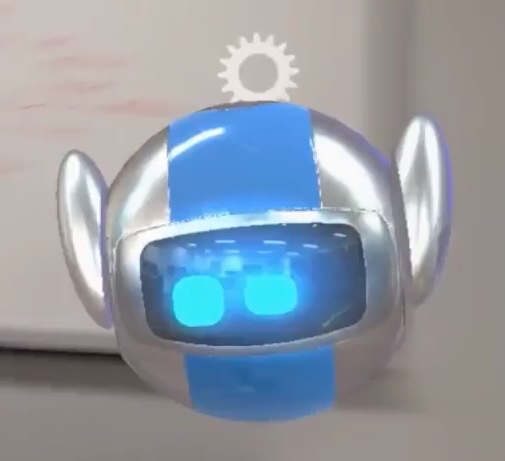}
    \caption{Cobot exhibiting a social cue when it is ``thinking'' (i.e. processing LLM prompts) by tilting its head. Cartoon-like cogs emanate from its head as a secondary cue.}
    \label{fig:cobot-thinking}
\end{figure}

\subsubsection{Listening/Thinking}
In listening/thinking state, Cobot moves in front of the user and provides eye contact as a social \textbf{attention}~\markemso{2} cue, orienting toward the user. Cobot provides head movements and emotes as social collaborative cues of \textbf{expression} \markemso{1}. As an example, shown in \autoref{fig:cobot-thinking}, Cobot will tilt its head when it is ``thinking'' (i.e. processing the LLM prompt). Such an anthropomorphic cue reduces user frustration and potential acceptance of the agent as a collaborator rather than a tool. 

\subsubsection{Active}
Finally, in active state, Cobot is free to navigate from point to point based on the actions generated by the LLM (the actions specified in the Lua code), under the constraints specified below:

\begin{enumerate}[leftmargin=*]
    \item \textbf{Trajectory:} When the Lua code requests Cobot to go to a new location, the system first calculates a straight line trajectory, unless the line overlaps with the user's personal space radius. Should this be the case, the trajectory is changed to a Bezier curve, using Cobot's current heading as start tangent and adjusting the end tangent to carefully avoid the user's personal space (enforcing \textbf{proxemics }\markemso{6}).
    
    \item \textbf{Avoid collision with the user at all costs:} Because the user is allowed to walk freely through the virtual space, it is possible for the user to walk into Cobot's path. To prevent this, we've added a repulsive force between the user and Cobot, the magnitude of which is inversely proportional to the squared distance between the two. When Cobot gets too close to the user, the force gets applied and a new path is calculated using the trajectory rules described above.
    
    \item \textbf{Physically-based acceleration:} To help the user predict Cobot's movement, Cobot's speed changes mimics those of a flying object, controlling a thrust (acceleration) vector in order to adjust its speed. This behaviour reinforces \textbf{physical coherence }\markemsp{3}.
\end{enumerate}

\begin{figure}
    \centering
    \includegraphics[width=0.75\linewidth]{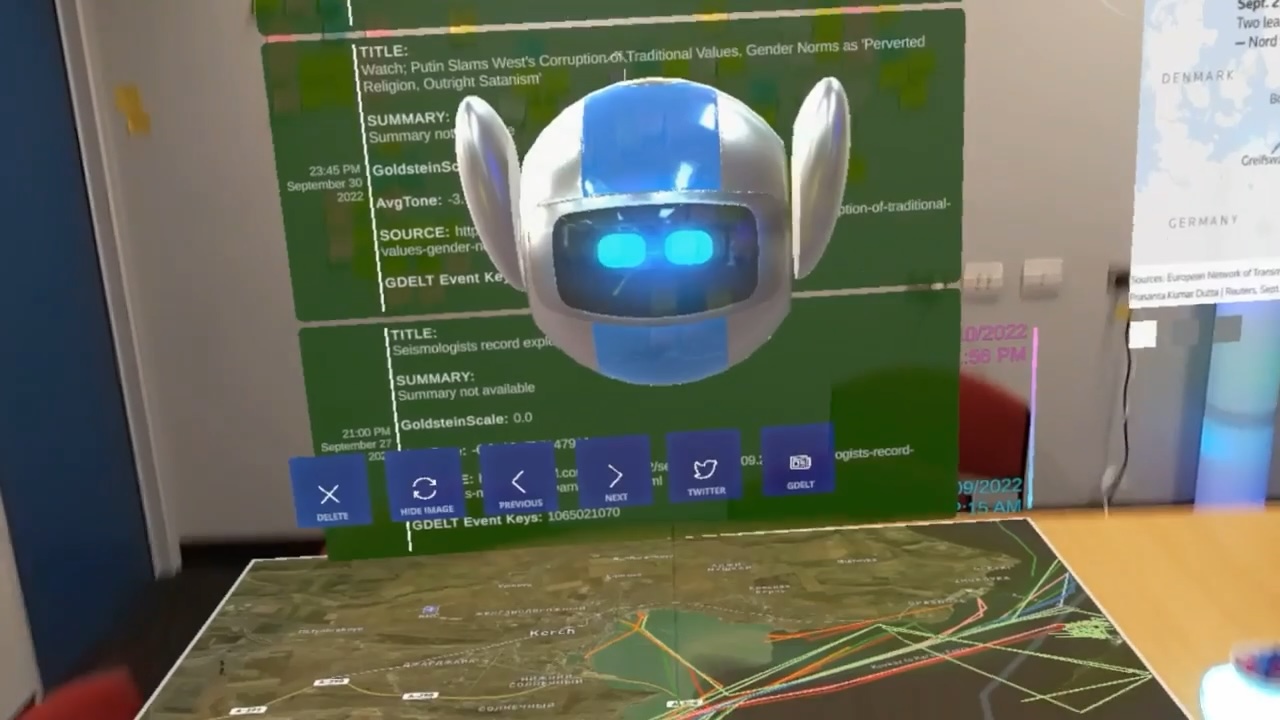}
    \caption{Cobot is spatially aware of the user's position and the visualisations in the environment.}
    \label{fig:cobot-spatial}
\end{figure}

Using these rules, Cobot will navigate move towards locations in the space where visualisation are to operate on them (\autoref{fig:cobot-spatial}), or to where they will be to create them, enabling semantic \textbf{proximity} \markemsm{2}. These rules produce a smooth, predictable movement, and keep Cobot from getting too close to the user and stepping into their personal space.

%% file: sections/usecase.tex
\section{Use Case Scenarios}
\label{sec:usecase}
\begin{figure}
    \centering
    \includegraphics[width=1\linewidth]{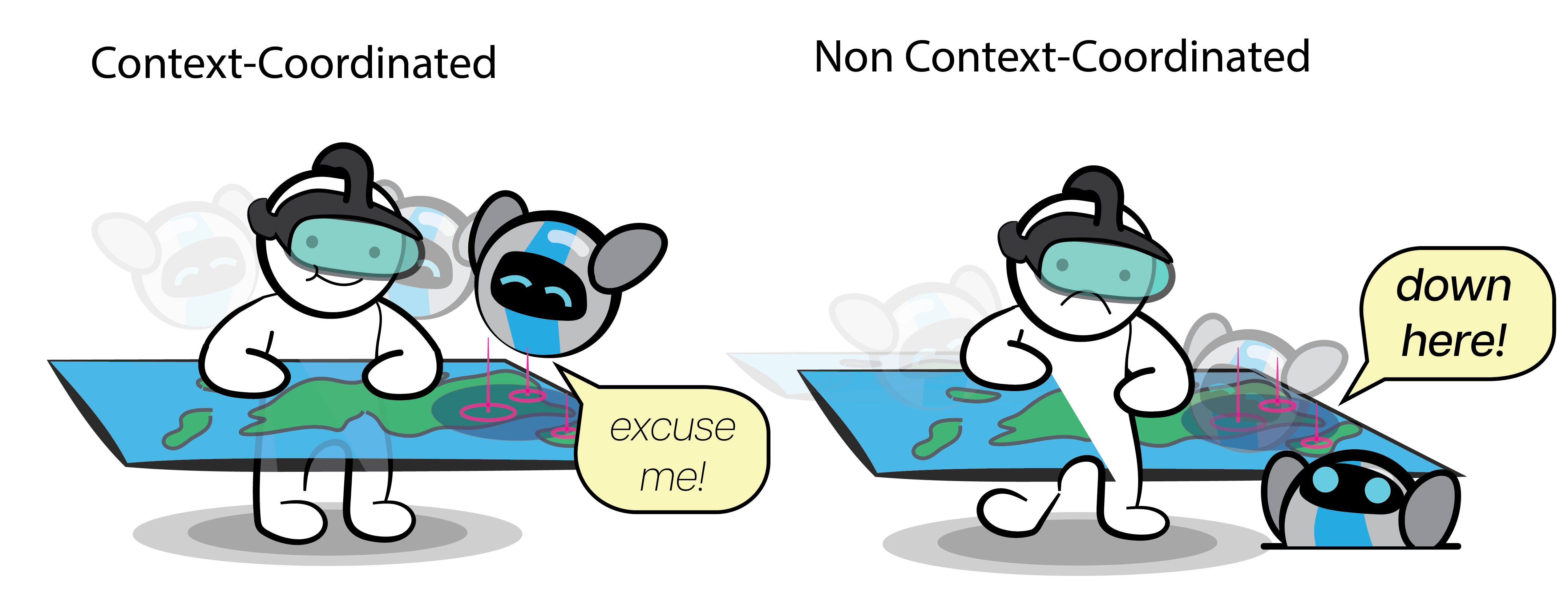}
    \caption{Comparisons of a \textit{context} and \textit{non-context} coordinated agent in an immersive analytics environment.}
    \label{fig:context-compare}
\end{figure}

In this section, we contrast the Cobot design probe against an AI agent that has not considered context across a use case scenario. In this scenario, ``\textit{Sophia}'' is a data analyst exploring a dataset of tracks from shipping vessels to understand patterns, trends and outliers.

We begin by describing an interaction with Cobot as a context-coordinated AI agent (illustrated in  \autoref{fig:context-compare}):
\begin{tcolorbox}[breakable,title=Context-coordinated AI agent]
Sofia is standing at her workspace, wearing an AR headset. Cobot hovers nearby, slightly behind her right shoulder not in the way, but present.

Sophia turns her head to look at Cobot and says, ``Show me global shipping activity this week.'' Cobot thinks for a minute and moves around and in-front of Sophia. Cobot projects a map of the world with all shipping data rendered as paths on the map. Cobot returns to Sophia’s side.

Sofia leans in to inspect: ``Cobot, Zoom in on that dense cluster near Singapore.''
Cobot moves towards the map again, zooming into selected region. Individual vessel tracks emerge.

Sofia asks, ``Cobot, what’s going on with that one?''
Cobot says ``That’s the MV Horizon Dawn. It’s been stationary for 14 hours.''

Sofia looks at Cobot and asks, ``List the events with that ship.'' Cobot thinks by tilting its head,  navigates to a clear part of the room, and a mini-timeline of events appears in-front of Cobot.

As Sofia walks around to view from another angle, Cobot smoothly repositions, never occluding her view or drifting through her body.

\end{tcolorbox}

In the following callout, we describe an embodied AI agent that is the antithesis to our conceptual framework (illustrated in \autoref{fig:context-compare}, \textit{right}). 

\begin{tcolorbox}[title=Non context-coordinated agent]

Sofia activates the system. The embodied AI materialises directly in front of her, clipping through Sophia and making her feel uncomfortable. 
Sophia moves backward to provide space and says, ``Show me global shipping activity.''
The agent pauses without feedback. 
After some time, Sophia turns and realises that the agent has created the map behind her, failing to be aware of the spatial context. This makes Sophia feel frustrated as she had been waiting for no reason.
She tries to inspect Singapore.

Sophia asks: ``Zoom in on Singapore.'' As the agent is unaware of context (e.g. user gaze), Sophia needs to clarify the particular visualisation to zoom in on, adding friction to the collaboration.

Lastly, Sophia asks: ``Show me all shipping events''. The agent teleports to another part of the room (failing to maintain physical coherence), surprising Sophia, and creates a timeline of events overlapping another visualisation.

\end{tcolorbox}

As we can see in this antithetical use case scenario, an agent that does not coordinate the spatial, social, and semantic contexts creates friction in the interactions. This friction can invoke negative feelings in the user, including frustration, surprise, and uncomfortableness. Furthermore, these forms of interaction likely increase cognitive load due to not matching the user's mental models and expectations, and may have negative impact on trust.2